\newcommand\showchange{TT}

\documentclass[conference]{IEEEtran}

\usepackage{amsmath,amssymb,amsfonts}
\usepackage{algorithmic}
\usepackage{graphicx}
\usepackage{textcomp}
\usepackage{xcolor}
\usepackage[hyphens]{url}

\usepackage[numbers,sort&compress]{natbib}

\usepackage[caption=false, font=footnotesize]{subfig}

\usepackage{tikz}
\usetikzlibrary{quantikz}

\usepackage{siunitx}
\usepackage{makecell}
\usepackage{booktabs}

\usepackage{hyperref}
\hypersetup{
    colorlinks=true,
    linkcolor=blue,
    filecolor=magenta,      
    urlcolor=cyan,
    citecolor=black,
    pdftitle={Overleaf Example},
    pdfpagemode=FullScreen,
    }

\usepackage{amsthm}
\usepackage{threeparttable}

\usepackage{multirow}

\usepackage{listings}
\usepackage[ruled,linesnumbered]{algorithm2e} 

\SetAlFnt{\small}
\SetAlCapFnt{\small}
\SetAlCapNameFnt{\small}
\SetAlCapHSkip{0pt}
\IncMargin{-\parindent}
\usepackage{soul}
\setstcolor{red}

\usepackage[normalem]{ulem}
\usepackage{cancel}
\newcommand\mycolrbox[3]{\colorbox{#1}{\parbox{.95\linewidth}{{\bf #2:} #3}}}
\long\def\com#1{}

\if\showchange

\newcommand{\TODO}[1]{\mycolrbox{yellow}{TODO}{#1}}
\newcommand{\FYI}[1]{\mycolrbox{lime}{FYI}{#1}}

\else

\newcommand{\TODO}[1]{}
\newcommand{\FYI}[1]{}
\fi

\newcommand{\blind}[1]{}        

\theoremstyle{definition}

\newcommand\name{{HiMA}\xspace}

\newcommand\myname {HiMA ({\bf Hi}erarchical {\bf M}icro-{\bf A}rchitecture)\xspace}

\newcommand\metrics{QPU load average\xspace}
\newcommand\metricsshort {QLA\xspace}

\newcommand\qubitsusage{qubits-per-circuit\xspace}
\newcommand\speedup{speedup\xspace}
\newcommand\speedupefficiency{speedup efficiency\xspace}

\newcommand\multithreading{quantum process-level parallelism\xspace}
\newcommand\multithreadingshort{quantum process-level parallelism\xspace}
\newcommand\distributed {hierarchical\xspace}

\newcommand\Value[0]{{\mathtt{v}}}

\def\BibTeX{{\rm B\kern-.05em{\sc i\kern-.025em b}\kern-.08em
    T\kern-.1667em\lower.7ex\hbox{E}\kern-.125emX}}

\title{HiMA: Hierarchical Quantum Microarchitecture for Qubit-Scaling and Quantum Process-Level Parallelism}
\author{
    \IEEEauthorblockN{
        Qi Zhou\texorpdfstring{\textsuperscript{1}}{}, 
        Zi-Hao Mei\texorpdfstring{\textsuperscript{1}}{},
        Han-Qing Shi\texorpdfstring{\textsuperscript{2}}{},
        Liang-Liang Guo\texorpdfstring{\textsuperscript{1}}{},
        Xiao-Yan Yang\texorpdfstring{\textsuperscript{1}}{},
        Yun-Jie Wang\texorpdfstring{\textsuperscript{1}}{},
        Xiao-Fan Xu\texorpdfstring{\textsuperscript{1}}{},\\
        Cheng Xue\texorpdfstring{\textsuperscript{4}}{},
        Wei-Cheng Kong\texorpdfstring{\textsuperscript{2}}{},
        Jun-Chao Wang\texorpdfstring{\textsuperscript{3}}{},
        Yu-Chun Wu\texorpdfstring{\textsuperscript{1}}{},
        Zhao-Yun Chen\texorpdfstring{\textsuperscript{4, $\dagger$}}{} and
        Guo-Ping Guo\texorpdfstring{\textsuperscript{1, 4, $\ddagger$}}{}
    }
    \IEEEauthorblockA{
        \textsuperscript{1}CAS Key Laboratory of Quantum Information, School of
        Physics, \\ University of Science and Technology of China, Hefei, Anhui, 230026, P. R. China\\
        \textsuperscript{2}Origin Quantum Computing Company Limited, Hefei, Anhui, P. R. China\\
        \textsuperscript{3}Laboratory for Advanced Computing and Intelligence Engineering, Zhengzhou, Henan, 450001, P. R. China\\
        \textsuperscript{4}Institute of Artificial Intelligence, Hefei Comprehensive National Science Center, Hefei, Anhui, 230088, P. R. China\\
        Email: $\dagger$chenzhaoyun@iai.ustc.edu.cn, $\ddagger$ gpguo@ustc.edu.cn
    }
}

\begin{document}

\maketitle\thispagestyle{plain}
\pagestyle{plain}


\begin{abstract}
Quantum computing holds immense potential for addressing a myriad of intricate challenges, which is significantly amplified when scaled to thousands of qubits. However, a major challenge lies in developing an efficient and scalable quantum control system. To address this, we propose a novel Hierarchical MicroArchitecture (HiMA) designed to facilitate qubit scaling and exploit quantum process-level parallelism. This microarchitecture is based on three core elements: (i) discrete qubit-level drive and readout, (ii) a process-based hierarchical trigger mechanism, and (iii) multiprocessing with a staggered triggering technique to enable efficient quantum process-level parallelism. We implement \name as a control system for a 72-qubit tunable superconducting quantum processing unit, serving a public quantum cloud computing platform, which is capable of expanding to 6144 qubits through three-layer cascading. In our benchmarking tests, \name achieves up to a 4.89× speedup under a 5-process parallel configuration. Consequently, to the best of our knowledge, we have achieved the highest CLOPS (Circuit Layer Operations Per Second), reaching up to 43,680, across all publicly available platforms.
\end{abstract}

\section{Introduction}
Recently, quantum computing has been shown to outperform classical computers in many computational tasks by state-of-the-art hardware with more than 100 qubits, including quantum supremacy~\cite{arute2019QuantumSupremacyUsing,wu2021StrongQuantumComputational,zlokapa2023BoundariesQuantumSupremacy} and quantum utility~\cite{kim2023EvidenceUtilityQuantum}. The escalation in the number of qubits is not only pivotal for realizing more practical quantum computing applications but also introduces significant challenges in control systems~\cite{bertels2020QuantumComputerArchitecture}. As a fundamental requirement, the quantum control system must be physically scalable and easily expandable. Furthermore, as the number of qubits increases, efficiently executing quantum programs and effectively utilizing quantum devices become critical issues. These requirements pose significant challenges to system scalability and underscore a crucial area for future research and development.

Centralized architectures~\cite{fu2017ExperimentalMicroarchitectureSuperconducting,fu2019EQASMExecutableQuantum,zhang2021ExploitingDifferentLevels}, which manage all qubit operations and parse quantum circuits through a single control unit, facilitate flexible feedback control~\cite{mallet2009SingleshotQubitReadout,walter2017RapidHighFidelitySingleShot,heinsoo2018RapidHighfidelityMultiplexed,montanaro2016QuantumAlgorithmsOverview,georgescu202025YearsQuantum,googlequantumai2023SuppressingQuantumErrors} and ease of compilation. However, the resource overhead for the control core, parsing efficiency, and the demand on I/O pin resources increase with the number of qubits, significantly limiting the scalability of this architecture. A natural progression is to decentralize quantum circuit information across various control units, each managing a subset of qubits, while still maintaining a central control core to implement system-wide feedback control~\cite{gupta2024encoding}. This approach mirrors the historical shift in classical computing from single-core to multi-core processors, motivated similarly by the need to enhance system scalability and processing power. 

\begin{figure}[htb]
	\centering
	\includegraphics[width=\linewidth]{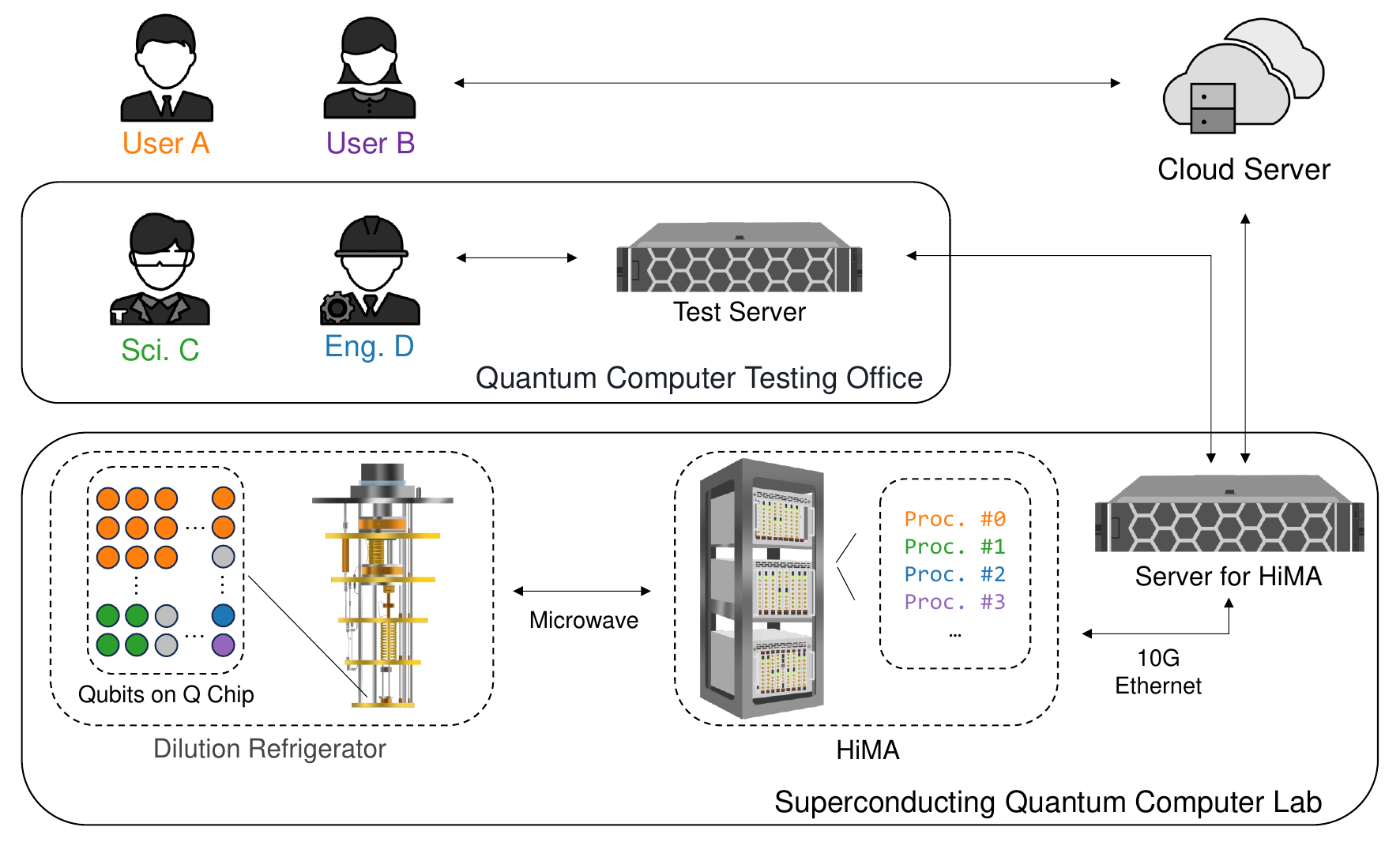}
	\caption{Scenario of Collaborative Quantum Computing. Multiple specialists concurrently test and develop applications on a shared quantum chip within a quantum computing laboratory. The system's process-level parallelism enables independent access to qubits by both on-site scientists and remote users via the cloud, fostering collaboration and accelerating quantum research.}
	\label{fig:background}
\end{figure}

Efficiency is intrinsically linked to the scalability of quantum control systems, as it directly influences how effectively resources are utilized as the system expands. Beyond enhancing the runtime proportion~\cite{smith2020OpensourceIndustrialstrengthOptimizing,gokhale2020OptimizedQuantumCompilation,cheng2020AccQOCAcceleratingQuantum}, another key aspect is improving the utilization of quantum device, e.g. providing quantum process level parallelism. 
Ref.~\cite{das2019CaseMultiProgrammingQuantum} proposes a software-level multiprogramming method, which merges compatible quantum circuits for execution. This approach increases parallelism to some extent but lacks flexibility. In contrast, hardware-level support for process-level parallelism introduces a more dynamic and efficient mechanism, allowing independent quantum processes to execute concurrently. 
This feature is particularly advantageous in scenarios where diverse tasks, such as qubit calibration experiments, are conducted simultaneously with the execution of quantum algorithms, significantly enhancing the system’s overall efficiency. As illustrated in Figure~\ref{fig:background}, the laboratory setting becomes a hub of activity where multiple specialists concurrently test and develop applications on a shared quantum chip. The system’s process-level parallelism is crucial for enabling independent access to qubits by both on-site scientists and remote users via the cloud. This collaborative framework not only fosters a more interactive research environment but also accelerates the pace of quantum advancements.

\begin{figure*}[htbp]
 \centering
 \footnotesize
 \centering
     \includegraphics[width=.8\linewidth]{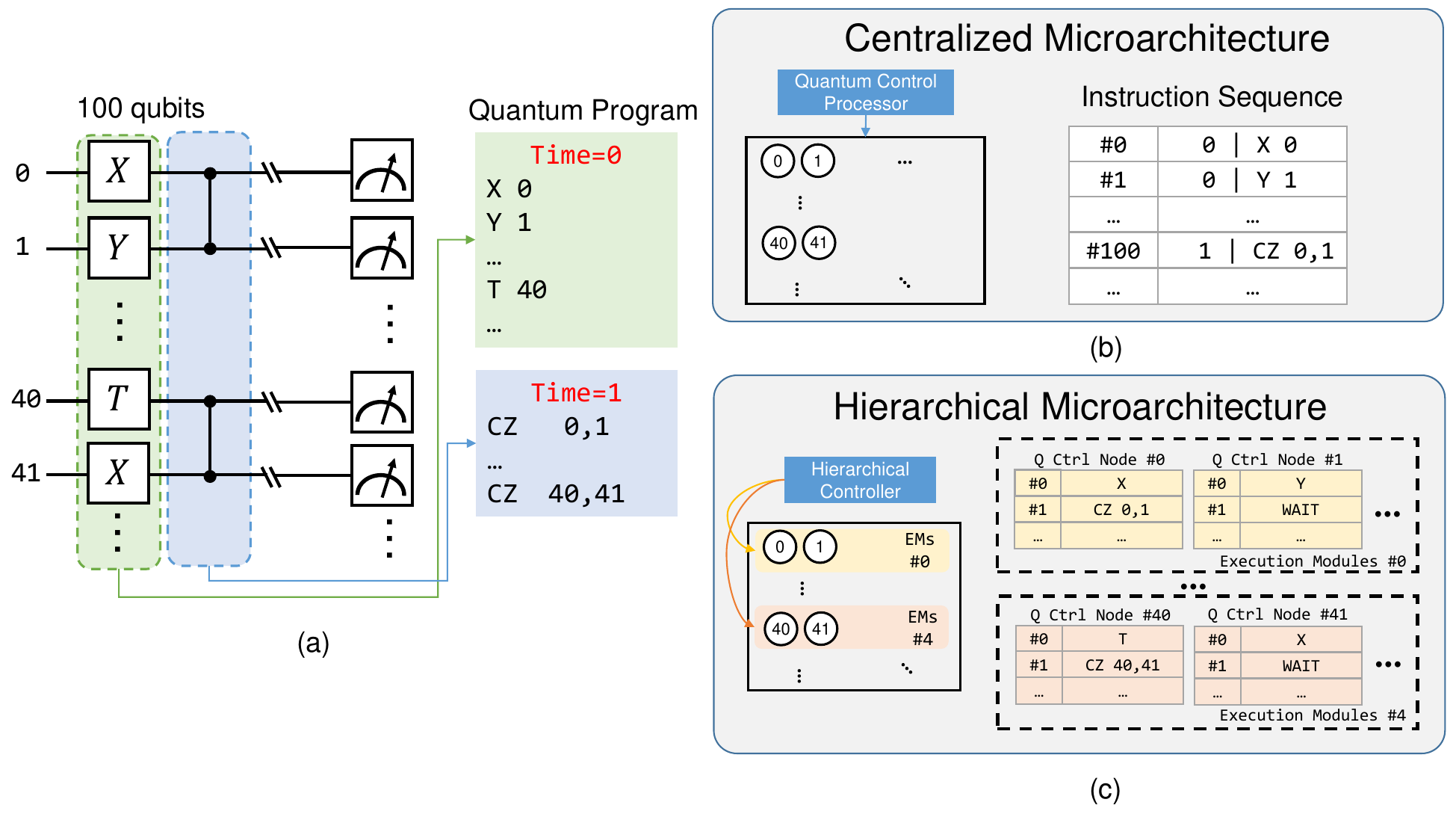}
     \caption{Comparison of hierarchical and centralized microarchitectures. (a) An example of a quantum circuit with 100 qubits. Each dashed box represents a layer of the quantum circuit, which should be executed in parallel. (b) A schematic diagram of how centralized microarchitecture handles the corresponding quantum program. For example, \texttt{0 | Y 1} indicates a Y-gate applied to qubit 1, occurring simultaneously with the previous instruction. Hence, the quantum control processor must parse a large number of instructions to allow the circuit-level parallelism. (c) A schematic diagram of the hierarchical microarchitecture. The quantum circuit is decomposed into quantum operation sequences for each qubit, which are executed within the corresponding qubit control nodes (QCNs). Synchronization is achieved through unified triggering by the root controller.}
     \label{fig:hierarchical} 
 \end{figure*}

In this paper, we introduce a novel control microarchitecture named \myname, designed to facilitate qubit scaling and quantum multiprocessing. Figure~\ref{fig:hierarchical} compares centralized and hierarchical microarchitectures. In a centralized architecture, a single quantum control processor is responsible for storing the entire quantum circuit using customized instructions. As the number of qubits increases, the corresponding growth in quantum circuit instructions leads to reduced execution efficiency and strained hardware resources, ultimately limiting scalability. In contrast, the hierarchical architecture distributes quantum circuit information across individual qubit control nodes (QCNs) associated with each qubit, with controllers synchronizing the timing of these units. This approach enhances scalability by cascading controllers without adding complexity or increasing the resource demands of the QCNs. Moreover, the controllers in \name support multiprocessing scheduling and management, enabling \multithreading and providing feedback control to accommodate complex quantum algorithms.

The main features and contributions of HiMA are:

\begin{enumerate}
    \item {\bf Discrete Qubit-Level Drive and Readout:} We utilize discrete execution units to individually store, parse, and execute XY and Z line drive and readout operations for each qubit. Notably, for qubit readout, we implement an asynchronous measurement method to enable control at the qubit level rather than the feedline level (measurement bus on QPU)~\cite{heinsoo2018RapidHighfidelityMultiplexed}. This strategy enhances system scalability through distributed storage. Furthermore, the independent control of each qubit allows for more flexible scheduling, establishing a robust foundation for quantum process-level parallelism.

    \item {\bf Process-Based Hierarchical Trigger Mechanism:} We employ cascading controllers to synchronize discrete execution units using a top-down, hierarchical trigger approach based on process numbers. This method facilitates efficient synchronization and scheduling of the execution units, minimizing resource overhead. 
    
    \item {\bf Multiprocessing Based Quantum Process-level Parallelism:} We adopt a multiprocessing approach to achieve quantum process-level parallelism. 
    To mitigate accuracy reduction induced by crosstalk between quantum processes, we deploy a staggered triggering strategy.
\end{enumerate}

Additionally, we implement \name as the quantum control system for a 72-qubit superconducting quantum device, which powers the publicly released Origin Quantum Cloud Platform.
HiMA can support up to 6144 fixed-frequency qubits through the introduction of three-layer cascading. To evaluate the system efficiency, we propose the \metrics (\metricsshort)  metrics that comprehensively evaluate the execution efficiency of quantum applications and utilization of QPUs. In a benchmark test, \name achieves up to 4.89$\times$ speedup under a 5-process parallel configuration.  Furthermore, we measure the system’s CLOPS through the cloud platform, and find that it can reach up to 43,680, with an eﬀiciency factor that is higher than the publicly available data from IBM and Rigetti. Finally, we validate our design by performing an interleaving randomized benchmarking (RB)~\cite{kayanuma2008CoherentDestructionTunneling,gambetta2012CharacterizationAddressabilitySimultaneous,proctor2017WhatRandomizedBenchmarking} experiment on a 72-qubit superconducting QPU, showing that \name can flexibly run multiple independent experiments in parallel properly.

\section{Background}
\label{sec:background}
\subsection{Quantum circuits}
The quantum circuit model, known as the most famous quantum computing model~\cite{jordan2008QuantumComputationCircuit}, is formed of quantum circuits that consist of quantum gates and measurement (a.k.a. readout)~\cite{nielsen2000QuantumComputationQuantum}. Before fully meeting a perfect quantum computer, we are currently in the Noisy Intermediate-Scale Quantum (NISQ) era~\cite{preskill2018QuantumComputingNISQ}, which adopts an imperfect quantum computer subject to noise and incoherence~\cite{harper2020EfficientLearningQuantum}. To seek applications in NISQ devices, two demands are needed: first, repeatedly executing the same quantum circuit to extract the probabilistic data on the quantum state, namely ``shots''~\cite{allahverdyan2013UnderstandingQuantumMeasurement}; second, performing mid-circuit measurement and feedback control to allow quantum error correction~\cite{georgescu202025YearsQuantum} and many quantum algorithms~\cite{montanaro2016QuantumAlgorithmsOverview}.

To enable quantum circuits to run on actual quantum computers, they need to be compiled to a particular quantum instruction set architecture~\cite{2017APS..MARP46008B}, taking into account microarchitectural and physical layer constraints. The compiled program will be translated into concrete operations of the quantum control processor.

Note that precise timing control, which is crucial for the implementation of two-qubit gates and some qubit experiments. To this end, the quantum control processor must ensure that issue rate is not less than the QPU execution rate. Several techniques~\cite{fu2019EQASMExecutableQuantum,zhang2021ExploitingDifferentLevels} are used to increase the issue rate and ensure timing.

\subsection{Superconducting qubits}
Quantum computers can be implemented using a variety of physical systems, among which superconducting computers are one of the most promising candidates to achieve both high integration and high fidelity~\cite{arute2019QuantumSupremacyUsing}. This paper mainly discusses quantum microarchitecture based on this physical system, while maintaining its compatibility with other systems.

To control superconducting qubits, one typically applies  microwave pulses, namely waveform sequence, on the XY line to implement single-qubit quantum gates~\cite{chen2016MeasuringSuppressingQuantum}. As for two-qubit gates (e.g. CZ, iSWAP)~\cite{foxen2020DemonstratingContinuousSet, yang2023ExperimentalImplementationShortPath}, the Z line of the target qubits and the tunable coupler (only for tunable qubits) between them need to be controlled. Typical execution time of single-qubit gates and two-qubit gates is on the order of \SI{10}{\nano\second}.

For qubit readout, a microwave pulse needs to be sent through the feedline to the readout cavity coupled to the qubit, then the transmitted signal from the readout cavity is collected~\cite{jeffrey2014FastAccurateState}. By analyzing the amplitude and phase of the collected signal, one can determine the measurement outcome being $\ket0$ or 
$\ket1$. To reduce the number of analog channels, the dispersive readout technique~\cite{walter2017RapidHighFidelitySingleShot} is commonly used, which mounts multiple readout cavities with different frequencies on a single feedline to achieve frequency multiplexing of measurement channels~\cite{heinsoo2018RapidHighfidelityMultiplexed}.

Note that qubits and quantum gates must be calibrated precisely before they are used for computation, requiring complex workflow described in~\cite{krantz2019QuantumEngineerGuide}. Due to the drift of quantum bit parameters during use, calibration experiments need to be performed regularly~\cite{kelly2014OptimalQuantumControl, kelly2018PhysicalQubitCalibration}.

\section{Defining Efficiency and Utilization of Quantum Processing Unit}
\begin{figure}[htb]
  \centering
  \includegraphics[width=\linewidth]{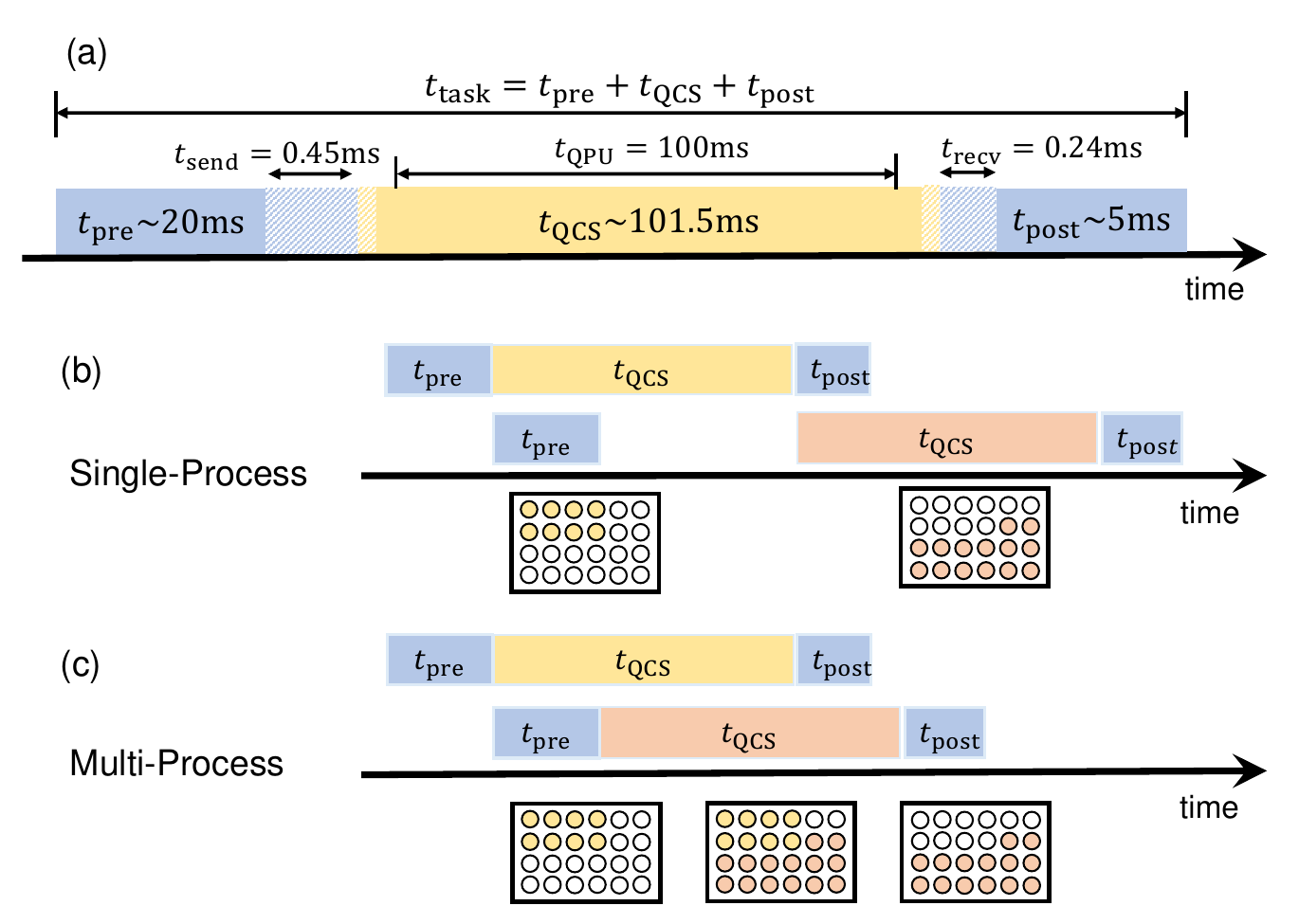} \\
  \caption{The time consumption span of the execution of a quantum application and schematics diagrams of the single- and multi-processing. (a) The total time for a quantum task can be divided into three parts: preprocessing ($t_{\rm pre}$), execution on quantum control system ($t_{\rm QCS}$), and postprocessing ($t_{\rm post}$). $t_{\rm send}$ and $t_{\rm recv}$ represent the time taken by the microarchitecture to receive data packets and send back results, which is short and will not be included in the following discussions. (b) In the single-process scheme, although the pre- and post-processing can be executed asynchronizingly, the time on the quantum chip is still the sum of two subsequent quantum programs. (c) In the multiprocessing scheme, when two quantum programs do not share the same qubit, they can be executed in parallel, causing an overlap in the time on the quantum chip.}
  \label{fig:timing_span} 
\end{figure}

\subsection{The time span and efficiency of quantum applications execution}

The execution process of quantum applications comprises the following three stages, as depicted in Fig.~\ref{fig:timing_span}. The preprocessing stage involves the reception and compilation of tasks. The execution phase, denoted by $t_{\rm QCS}$, primarily consists of $t_{\rm QPU}$, which represents the pure execution time on the quantum chip, i.e., the waveform play duration. In addition to $t_{\rm QPU}$, this phase also includes the time overhead for task parsing, distribution, loading and synchronization. However, these times are a minor part and will not be further discussed. In the postprocessing stage, the server performs advanced processing of readout results and transmits the processed data back to the client.

\newcommand{\descriptionwidth}{0.53\textwidth}
\renewcommand{\arraystretch}{1}

\begin{table*}[htb]
	\caption{Essential instructions of \name}
	\label{tab:QISA}
	\centering
	\begin{threeparttable}
		\begin{tabular}{llcl}
			\toprule
			\multicolumn{2}{c}{Syntax\tnote{1}} & Scope\tnote{2} & \multicolumn{1}{c}{Description} \\ 
			\midrule
			
			\multirow{1}{*}{\texttt{GATE}} & 
			\multirow{1}{*}{\texttt{addr, dur, trig}}&
			\multirow{1}{*}{O} &
			\multirow{1}{\descriptionwidth}{Play the waveform from address \texttt{addr} for the duration specified by \texttt{dur}.} \\
			
			\texttt{WAIT} & \texttt{dur, trig} &
			\multirow{1}{*}{C, I, O} & 
			\multirow{1}{\descriptionwidth}{Wait for the duration specified by \texttt{dur}. Used for timing alignment.} \\
			
			\texttt{MEASURE} \tnote{3} & \texttt{dur, dtype, fb, trig} &
			\multirow{1}{*}I &
			\multirow{1}{\descriptionwidth}{Acquire data for the duration specified by \texttt{dur}.}\\
			
			\texttt{TRIGGER} & \texttt{start, trig} &
			\multirow{1}{*}C & 
			\multirow{1}{\descriptionwidth}{Send a synchronized trigger signal to exection modules.} \\ 
			
			\multirow{1}{*}{\texttt{FEEDBACK}} &
			\multirow{1}{*}{\texttt{addr}} &
			\multirow{1}{*}{C} & 
			\multirow{1}{\descriptionwidth}{Perform feedback interrupt based on the memory space specified by \texttt{addr}} \\
			
			\texttt{BR} & \texttt{rs, imm ,offset} &
			\multirow{1}{*}{C, I, O} & 
			\multirow{1}{\descriptionwidth}{Jump to \texttt{PC + Offset} if \texttt{rs} is equate to \texttt{imm}. \texttt{rs} stores feedback result.} \\
			\bottomrule
		\end{tabular}
		\begin{tablenotes}
			\item[1] \verb|dur| stands for duration. \verb|trig| stands for trigger flag as mentioned in Section~\ref{sec:microarchitecture}. \verb|start| stands for start flag as mentioned in Section~\ref{sec_QPLP}.
			\item[2] ``C", ``I", ``O" represents the controller, the qubit readout input unit, and two components: the XY/Z drive unit and the qubit readout output unit, respectively.
			\item[3] \verb|dtype| stands for output data type. 0, 1, 2 represent for 
			qubit state, intermediate results and original input data, respectively. \verb|fb| stands for feedback flag. When feedback flag is set to 1, the readout results are forwarded to the controller for feedback control.
		\end{tablenotes}
	\end{threeparttable}
\end{table*}

As only a fraction of the execution time for a quantum application is spent utilizing the computational resources of the QPU, the execution efficiency of quantum application can be calculated as $t_{\text{QPU}}/t_{\text{total}}$. Prior research has mainly focused on reducing $t_{\text{total}}$ by minimizing the amount of transmitted data, improving instruction parsing speed~\cite{batabyal2019RealizingParallelismQuantum} and preprocessing time span~\cite{heckey2015CompilerManagementCommunication}, consequently enhancing efficiency.

\subsection{Quantum Process-level Parallelism and QPU Load Average}

In this paper, we seek another way to improve the efficiency by increasing the utilization rate in terms of qubits. Luckily, the control channels of superconducting qubits demonstrate a notable degree of independence, as detailed in Section~\ref{sec:background}, which naturally results in a parallelism in scheduling QPU resources. That is, if multiple processes do not conflict in qubits and channels, they can be executed in parallel~\cite{mineh2023AcceleratingVariationalQuantum, das2019CaseMultiProgrammingQuantum}, which is called \textbf{``Quantum Process-Level Parallelism''}. For example, in Fig.~\ref{fig:timing_span}(b), Process 1 uses the yellow region of the quantum chip while Process 2 uses the red region. When process-level parallelism is not allowed, the total execution time on the QPU is the sum of these two tasks. Here, our target is to allow these two processes to be executed in parallel, greatly reduces the time cost due to the overlap, which is shown in Fig.~\ref{fig:timing_span}(c).

A comparable strategy can be achieved through compile-time merging
of tasks that can be parallelized within a time slice~\cite{das2019CaseMultiProgrammingQuantum}. However, this merging and compilation approach brings additional time overhead. Additionally, the issue of the QCS blocking other tasks while executing a task remains unresolved. Moreover, the efficiency of the merging and compilation strategy is directly dependent on the granularity of the time slice and is closely related to the duration of the executed quantum circuits. Therefore, native support for quantum process-level parallelism would be significantly more flexible and efficient.

In the following two typical scenarios, hardware support for quantum process-level parallelism can significantly enhance quantum device utilization, particularly when managing a large number of qubits.

\begin{itemize}
    \item Qubits are inherently susceptible to noise fluctuations, necessitating periodic calibration to ensure the accuracy of computational results. During the execution of quantum algorithms, it may be necessary to perform calibration procedures on qubits located in other regions of the device. Supporting quantum process-level parallelism can effectively address this complex scenario by independently scheduling and managing different regions of the quantum device. 

    \item In the realm of quantum cloud computing, both the number of qubits required and the timing of user-submitted quantum applications are inherently uncertain. Quantum process-level parallelism enables the asynchronous execution of quantum circuits, which allows for more flexible scheduling of computational resources on the quantum device. This flexibility significantly enhances the efficiency of execution, optimizing resource utilization and accommodating the dynamic needs of different users efficiently.
    
\end{itemize}

To characterize the efficiency of the task execution, showing how the system benefits from the process-level parallel in the context of processing continuous tasks, we propose the concept of the \metrics, defined as
\begin{equation}
    \mathrm{QLA} = \frac{\sum t_{QPU_i} \times n_i}{t_{total} \times N},
    \label{eq:QLA_}
\end{equation}
where $n_i$ represents the number of qubits used in the $i_{th}$ task, $t_{QPU_{i}}$ is the execution time of the $i_{th}$ task on the QPU, $N$ denotes the total number of qubits in the QPU, and $T_{total}$ is the total execution time for a series of tasks.

\subsection{Circuit
Layer Operations Per Second}

CLOPS~\cite{wack2021quality}, proposed by the IBM Quantum team, is a comprehensive measure of the operational speed and the quality of the quantum system. The test of CLOPS emulates the variational quantum algorithm scenario: it sets several instances of parametric quantum circuits, where the parameter of each circuit is iteratively updated according to the readout results of the last circuit. Then the CLOPS can be calculated as $(M\times K\times S\times D)/\mathrm{Time}$, where $M$, $K$, $S$ and $\mathrm{Time}$ represents the number of instances, number of iterations, number of shots and execution time of the whole process, respectively. In the standard CLOPS test, $M=100,\,K=10,\,S=100$. $D=\log_2 \mathrm{QV}$ is the logarithm of quantum volume, so that CLOPS also reflects the quality of the quantum processor (quantum volume), which is unrelated to this paper. To merely characterize the control system's efficiency, we define the \textbf{efficiency factor} from dividing CLOPS by QV, which only corresponds to the time of finishing a CLOPS test.

\section{Requirement}

This section introduces specific requirements for microarchitecture of quantum control system. 

\subsection{Scalability}

Scalability is a primary consideration in the design of control microarchitectures for quantum control systems, necessitating conditions that ensure the system can expand efficiently without compromising performance. Two main factors restrict scalability:

\begin{itemize}
    \item {\bf Resource Overhead of the Control Core:} Current microarchitectures often employ field-programmable gate arrays (FPGAs) as controllers. The limited computational, I/O and memory resources of FPGAs can hinder scalability. A control core is essential for a quantum control system used for feedback control, synchronization and communicating with execution modules. Therefore, it is vital for the resource demands on the control core to grow minimally relative to the increase in the number of qubits, while also ensuring that the core’s resources can be expanded.
        
    \item {\bf Issue Rate:} Accurate outcomes for quantum applications require that the real-time parsing speed of quantum circuits exceeds the execution time of quantum gates. This requirement implies that the time overhead associated with quantum circuit parsing should not increase with the number of qubits.
    
\end{itemize}

\subsection{Timing Synchronization}
Timing synchronization plays a vital role in quantum control systems, as the precise sequence of quantum operations significantly affects the reliability of results. Each quantum operation on individual qubits, as well as the coordination among operations across different qubits, needs to adhere to a pre-scheduled timeline to maintain the accuracy of quantum applications. Additionally, to support effective quantum process-level parallelism, the architecture needs to ensure not only the consistency of timing triggers within a single task but also the independence of timing between concurrent tasks.

\subsection{Feedback Control}

Feedback control is essential in quantum computing, particularly for applications like qubit fast reset and quantum error correction. To implement effective feedback control, several conditions are essential:

\begin{itemize}
    \item The architecture should support the real-time updating of quantum circuit parsing based on feedback results.
    \item The control core should determine the direction of feedback data for each qubit, based on readout results.
    \item High-speed and low-latency data transmission links are essential for realizing efficient feedback loops. 
    
\end{itemize}

\section{Microarchitecture} \label{sec:microarchitecture}

\subsection{Overview of \name}
\label{sec:microarchitecture_structure}

\begin{figure*}[htb]
  \centering
  \includegraphics[width=.95\textwidth]{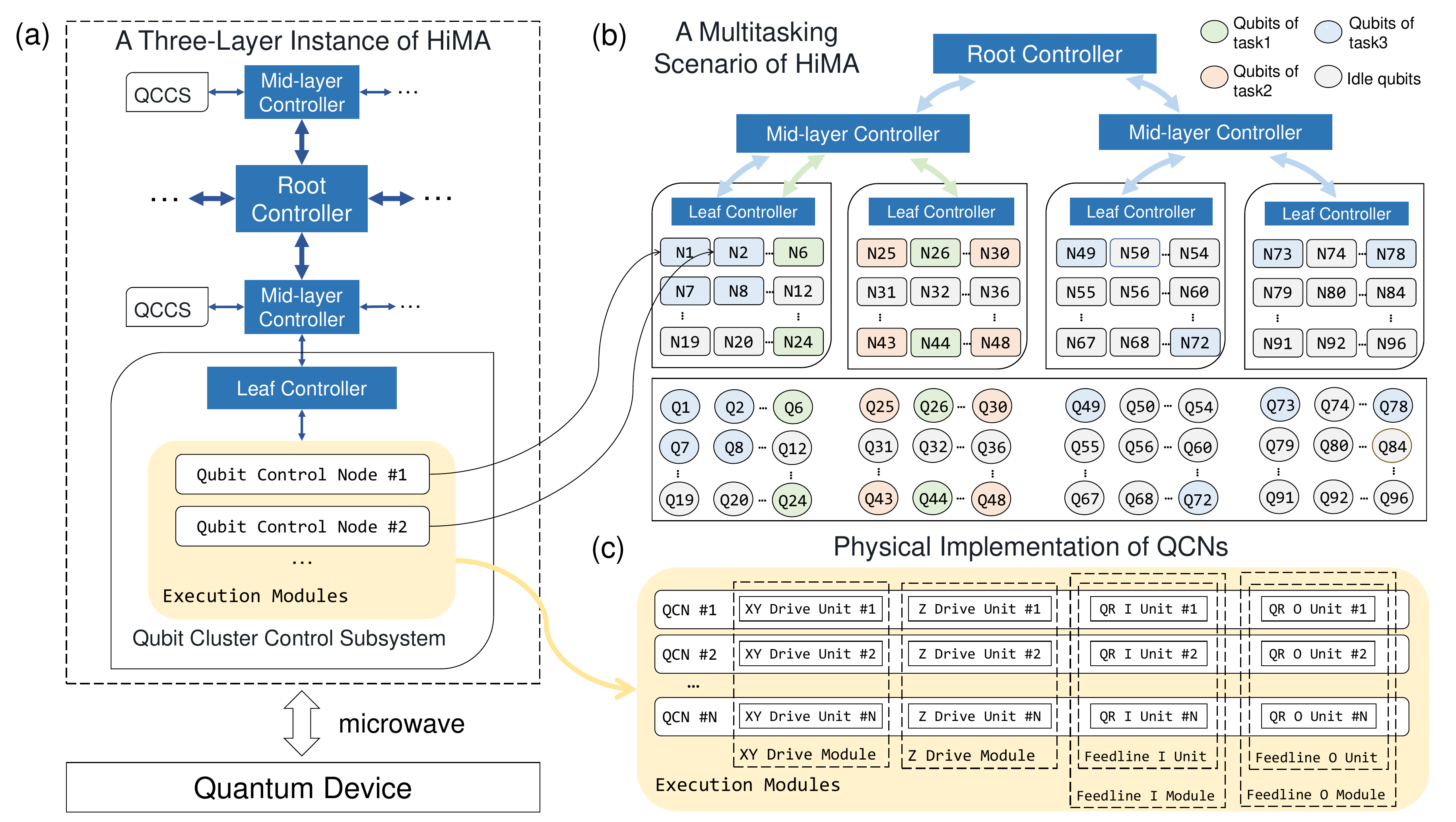}
  \caption{
    (a) A three-layer instance of HiMA, demonstrating its cascade structure. The root controller interfaces through middle-layer controllers, which are connected to multiple qubit cluster control subsystems (QCCSs). Each subsystem includes a leaf controller and qubit control nodes (QCNs) housed within execution modules.
    (b) A Multitasking Scenario of HiMA. Each QCCS is abstracted as consisting of 1 leaf controller and 24 QCNs, depicted as a diagonal corner rounded rectangular box. HiMA supports the asynchronous parallel execution of multiple tasks, with each task involving different QCNs.
    (c) Physical implementation of QCNs within execution modules. Execution modules include XY/Z drive modules and feedline input/output (I/O) modules. Each XY/Z drive module contains multiple qubit-level XY/Z drive units, while each feedline I/O module includes several feedline I/O units, with each unit housing multiple qubit readout (QR) I/O units corresponding to the number of qubits per feedline. A QCN is constructed by integrating the relevant qubit-level units from these modules.
  }
  \label{fig:dis_micro}
\end{figure*}

\subsubsection{System overview}
In this paper, we propose a \distributed microarchitecture to achieve \multithreading. Essential instructions of \name are shown in Table~\ref{tab:QISA}. As illustrated in Fig.~\ref{fig:dis_micro}(a), \name is composed of the \textbf{controllers} and {\bf execution modules}. The execution module, comprising {\bf XY/Z drive modules} and {\bf feedline input/output (I/O) modules}, is responsible for qubits drive and readout, storing waveform and timing information of the quantum circuit. Additionally, it supports executing subsequent quantum circuits based on feedback data. The controller's primary role is to synchronize the timing between the execution modules and make feedback decisions. The runtime interaction between the execution module and the controller involves two types. Trigger links are employed to maintain the timing consistency between execution modules. Feedback links are used for transmitting the readout results of qubits and feedback data.

The hierarchical architecture of HiMA is manifested both physically and functionally. To address the issue of limited external interfaces and computing resources in realistic hardware, a cascading controller architecture, known as the ``hierarchical architecture'', is introduced to ensure the physical scalability of HiMA.  Fig.~\ref{fig:dis_micro}(a) shows a three-level cascaded system. A leaf controller, along with all the execution modules it controls, constitutes a {\bf qubit cluster control subsystem (QCCS)}. The {\bf root controller}, mid-layer controller and multiple QCCSs are arranged in a starlike topology, ensuring the scalability of our architecture.

Functionally, the implementation of discrete qubit-level drive and readout significantly enhances the system's flexibility, serving as the foundation for multiprocessing. Fig.~\ref{fig:dis_micro}(c) illustrates the qubit readout input/output and XY/Z drive unit of the same qubit, abstracted as the {\bf qubit control node} (QCN) on the server side. This design effectively decouples the physical area associated with qubit-level control and channels within the microarchitecture. Execution units involved in the same task can synchronize timing through the {\bf task control processor} located in the controller. Moreover, the controller is equipped with multiple task control processors that are capable of managing multiple tasks independently and in parallel. Consequently, HiMA is capable of executing a large-scale quantum circuit comprehensively, as well as performing asynchronous parallel execution of various quantum circuits or qubit calibrations. Fig.~\ref{fig:dis_micro}(b) displays HiMA executing three independent quantum circuits asynchronously. Task2 only requires execution within a single QCCS, while task 1 and task 3 require synchronization and feedback between QCCSs through the mid-layer and root controller.

\subsection{Discrete Qubit-Level Drive and Readout}
The execution module consists of XY/Z drive modules and feedline I/O modules,  which are responsible for qubits drive and readout, respectively. Discrete qubit-level drive and readout refer to storing, parsing and executing the XY/Z drive and readout operations for each qubit independently through different execution units. Hence, the time overhead of quantum circuit parsing and storage resources do not increase with the increase of qubit numbers, greatly enhancing scalability. This approach is also conducive to the flexible scheduling of each execution unit.

\subsubsection{Qubit Drive Unit}

\begin{figure}[htbp]
\centering
\footnotesize
\centering
    \subfloat[]{
        \includegraphics[width=0.45\textwidth]{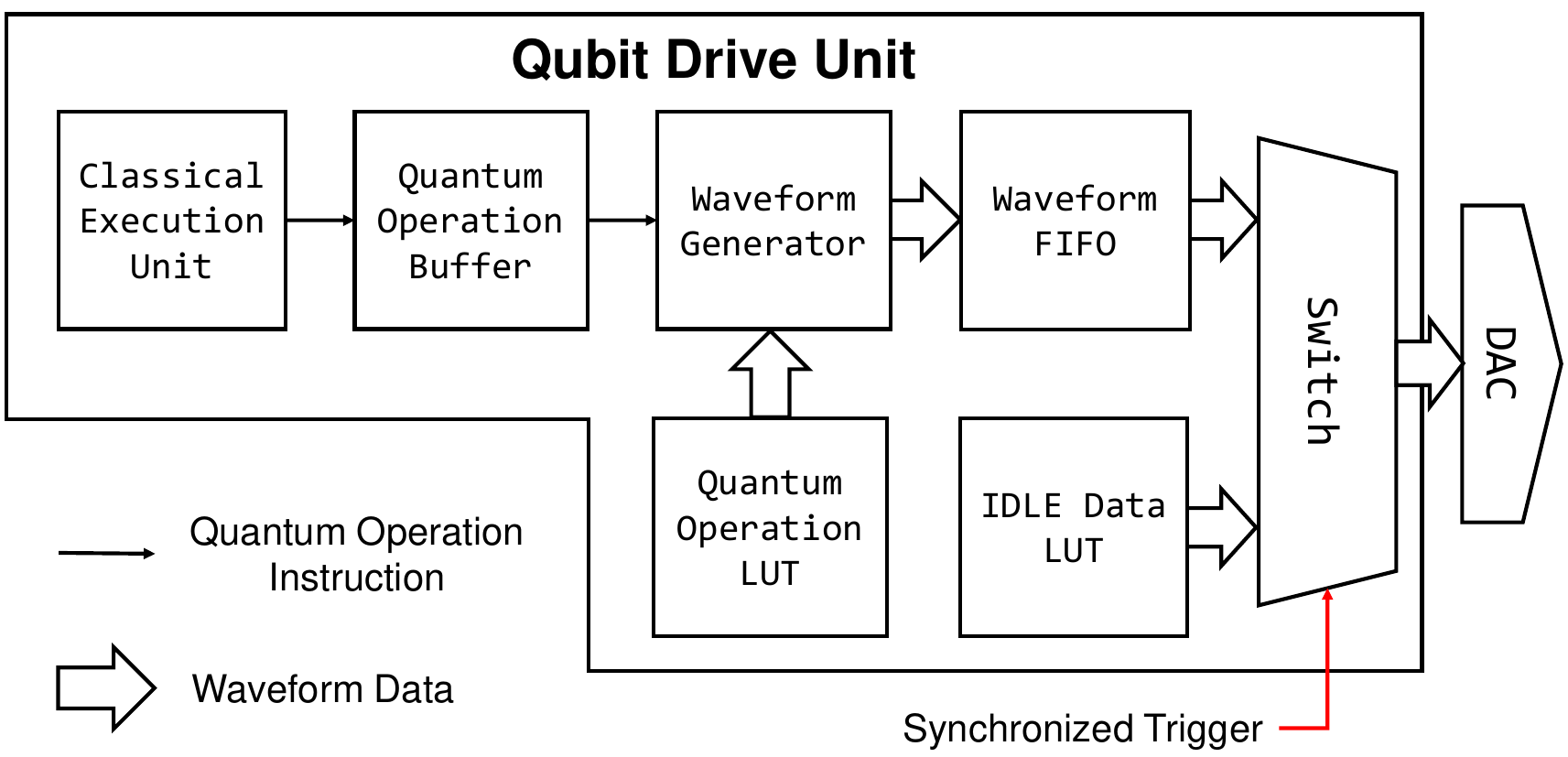}
    }\hfill
    \subfloat[]{
        \includegraphics[width=0.45\textwidth]{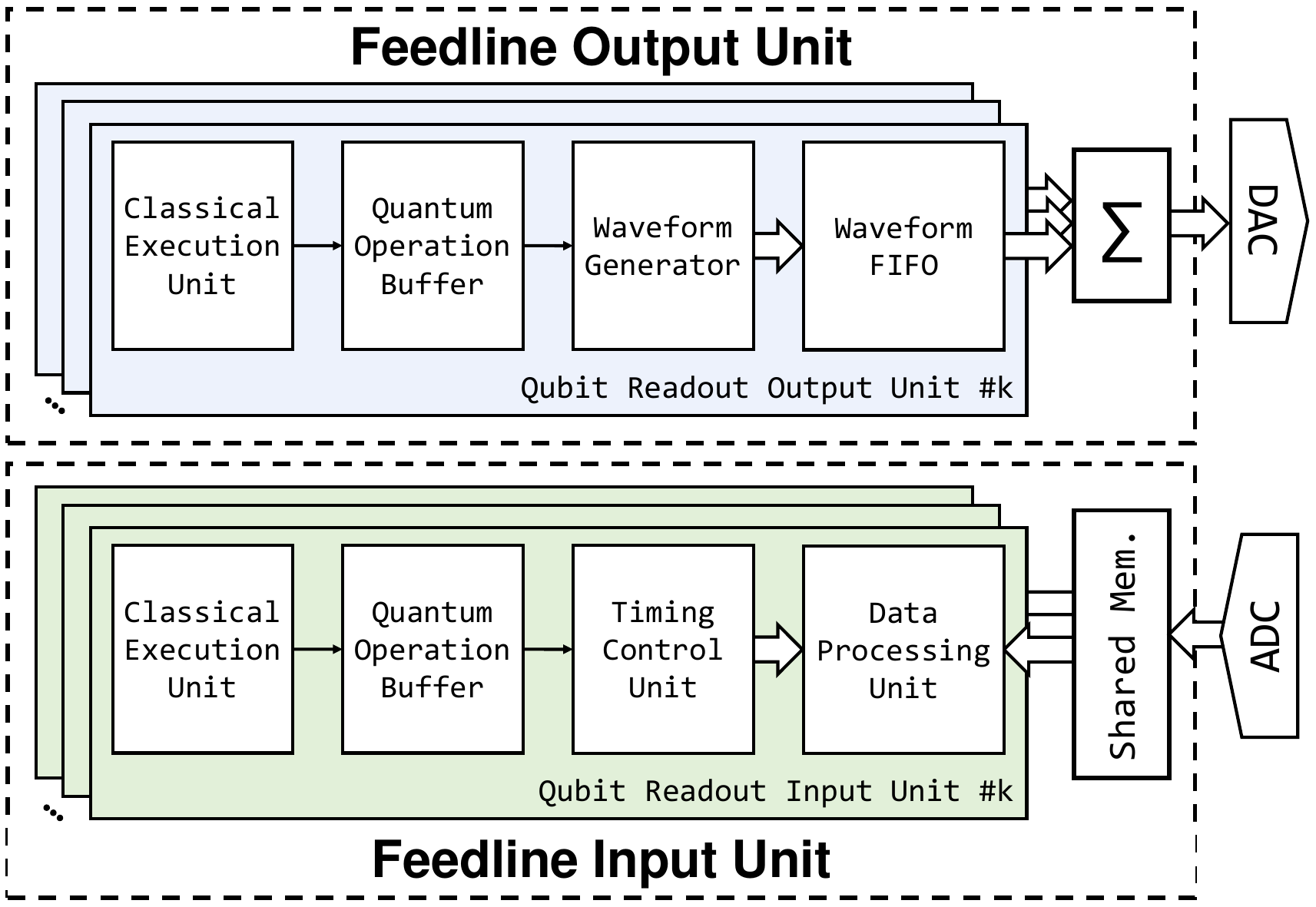}
    }
    \caption{(a) Architecture of the qubit drive unit within the XY/Z drive module. (b) The feedline Input/Output (I/O) unit manages the feedline, while the qubit readout I/O units handle the readout operations for the qubits sharing that feedline.}
    \label{fig:qubitcontrol} 
\end{figure}

Each output channel for XY and Z drive of qubits is independently controlled by a {\bf qubit drive unit}. This unit stores quantum operations sequences of corresponding qubit and converts them into waveform sequences for qubit drive through digital-analog converters (DACs). The architecture of the  qubit drive unit is depicted on the left side of Fig.~\ref{fig:qubitcontrol}.

The classical execution unit is primarily employed for parsing quantum circuits and feedback control, as further explained in Section~\ref{feedback}. It sends quantum operation instructions into quantum operation execution unit, decodes them into sets of waveform sequences according to the quantum operation look-up table (LUT), and writes into the waveform first-in, first-out (FIFO) buffer. The waveform FIFO serves as a vital interface between the instruction execution and QPU timing domains. The waveform generator converts quantum operations into waveform sequence that can be directly used by the DAC chip, and sequentially writes into the waveform FIFO. We configure the read and write rates of the FIFOs to be equal to the DAC output rates. With this setup, as long as the data is continuously read from the waveform FIFO, the timing of quantum operations output by the DAC is assumed to be in strict accordance with the quantum circuit as described in the instructions. To ensure the efficiency of the waveform FIFO is not blocked by classical instructions, a quantum operation buffer is employed to decouple their execution from waveform generation. As the waveform generation module writes to the waveform FIFO, the classical processor can continue to send quantum operation instructions until the quantum operation buffer reaches its full capacity, or an interrupt is raised by the execution of a \verb|BR| instruction. 

Since the output channels are physically separated, it is essential for the drive module to synchronize the runtime of the QPU across multiple channels. Here, we use a common trigger signal from the cascaded controller to synchronize the execution module. The quantum operation instruction is defined as \verb|GATE dur, trig|, with \verb|dur| indicating the duration of the operation, and \verb|trig| is used to trigger synchronization. The output selector will hold waveform sequences with a trigger flag position of 1 and send the waveform of IDLE operation to the DAC chip instead until it receives a synchronized trigger signal.

\subsubsection{Feedline Input/Output Unit}

Unlike the XY or Z drive of individual qubits, multiple qubits share a single feedline used for qubit readout. We propose an asynchronous measurement scheme to ensure discrete qubit readout operations. Fig.~\ref{fig:qubitcontrol}(b) illustrates the design of the feedline input/output (I/O) unit.

The qubit readout I/O unit, as the core component of the feedline I/O unit, independently stores data concerning the measurement operations and quantum state discrimination waveforms of the corresponding qubits. The processes of timing control and feedback are implemented similarly to those in the drive module and will not be discussed in detail here.

The waveform sequence for the feedline is generated in real-time using frequency division multiplexing. The outputs from qubit readout output units that share the same feedline are sent to a multi-input adder to generate a real-time waveform sequence for the measurement operation of the feedline. Regarding data input, we continuously collect data from each feedline, store it in shared memory, and broadcast it to all related qubit readout input units. When the \verb|MEASURE| instruction is executed, the qubit readout input unit processes the input data in real-time and obtains the readout result.

\subsection{Process-Based Hierarchical Trigger Mechanism}

After implementing discrete drive and readout at the qubit level, establishing timing synchronization becomes essential. This includes synchronizing the execution units of the same qubit and aligning the timing of different qubits within the same quantum circuit. With qubit-level control already mapped to the execution units during compilation, the primary task is to implement synchronization among these units. We employ a synchronous trigger mechanism for this purpose. The root controller initiates synchronization by sending synchronous trigger signals in a stepwise fashion down the hierarchy. The moment the execution module receives this trigger signal marks the beginning of a timing period, which could represent the start of a shot or the initiation of a quantum circuit following feedback. Effective synchronization is achieved by maintaining a consistent relative relationship between channels at each trigger, and ensuring that the timing within the execution module aligns precisely with the instructions.

\begin{figure*}[htbp]
\centering
\footnotesize
\centering
    \includegraphics[width=0.95\textwidth]{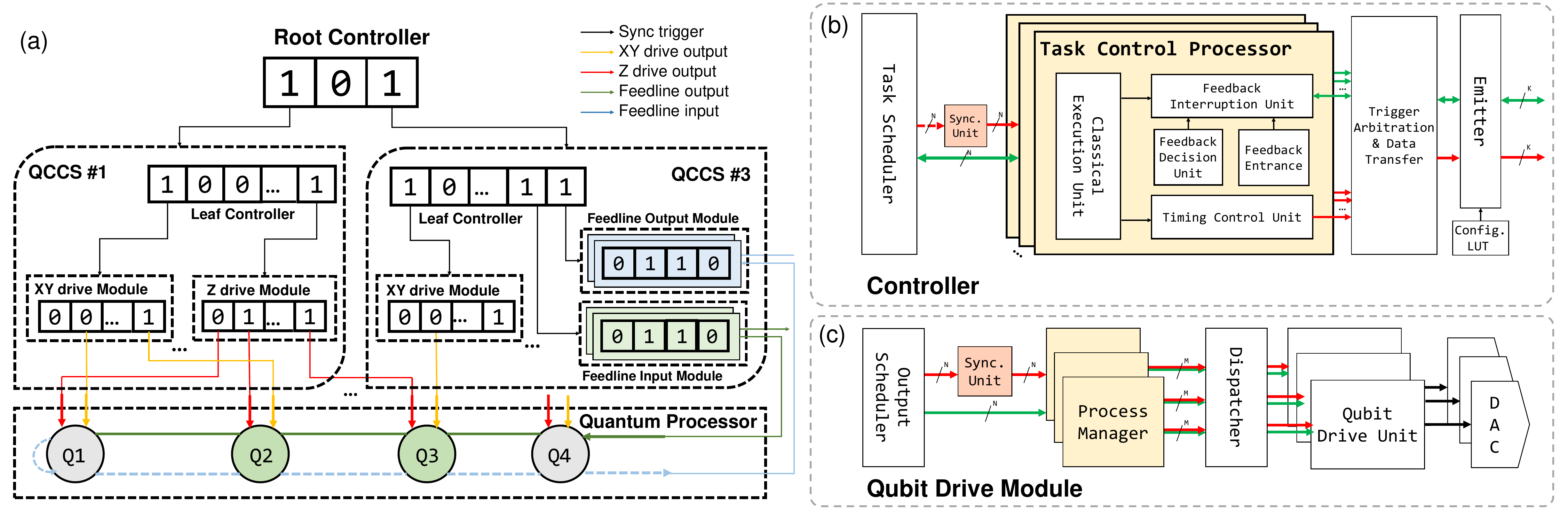}
    \caption{ (a) Demonstration of a process-based hierarchical trigger mechanism within a secondary cascade architecture: The task involves two qubits, Q2 and Q3, each controlled by distinct QCCSs. An arrow with the execution unit unconnected indicates that the channel is not connected to the current four qubits. Similarly, an arrow with Q4 unconnected indicates that this qubit is not controlled by the drive units shown in the figure. The root controller records the mask of each QCCS. Within each subsystem, the leaf controller records the mask of both the leaf controller and the feedline input/output (I/O) module corresponding to the task. Similarly, the execution module labels the mask of the corresponding execution unit, thus enabling top-down trigger triggering. (b) Multiprocessing based architecture of controller: This architecture features multiple task control processors that handle different quantum processes in parallel. (c) Drive module architecture: Multiple process managers oversee different processes. Each process manager is responsible for all qubit drive units.}
    \label{fig:multiprocess} 
\end{figure*}

\subsubsection{Controller}

The controller primarily focuses on ensuring synchronization trigger and making feedback decisions between the execution modules, which is a center core of quantum program execution.  

The {\bf task control processors} are the core part to implement functions of controller, as shown in Fig.~\ref{fig:multiprocess}(b). The classical execution unit handles program flow control for feedback. The trigger unit is utilized for executing instructions \verb|WAIT| and \verb|TRIGGER start trig|. For the initial trigger in each shot, \verb|start| is set to 1 to facilitate staggered triggering, especially in multiprocessing scenarios. For the middle-layer controllers, upon receiving the trigger signal from the root controller, the timing control unit initializes the runtime and triggers the execution modules. For the root controller, once all the execution and leaf controllers participating in the task are prepared, it initiates the timing and dispatches trigger signals to the leaf controllers. 

In addition, we adopt a synchronization protocol to ensure consistent relative channel relationships and phase synchronization among various qubits at each shot. The propagation of the trigger signal across physically distinct execution modules and controllers introduces a risk of metastability~\cite{ginosar2011MetastabilitySynchronizersTutorial}, potentially leading to discrepancies in the timing of trigger events across different modules. At system initialization, we generate a periodic synchronization signal with the same period across all modules. As our system operates on a synchronous clock, the execution modules, leaf controller and root controller maintain a fixed relationship with this periodic synchronization signal. Upon receiving the trigger signal from the controller, the execution module's synchronization unit waits for the arrival of the periodic synchronization signal before passing the trigger signal to the respective execution unit.

For achieving high fidelity in two-qubit gates, it is crucial that the phase difference in microwave pulses on the XY line, applied to different qubits, remains constant in each shot. The synchronous trigger mechanism ensures only the initial phase alignment of the intermediate frequency (IF) signal. We therefore set the periodic synchronization signal to an integer multiple of the periods of all local oscillators. This configuration ensures phase consistency among various qubits at the start of each shot.

\subsubsection{Process-Based Hierarchical Trigger Mechanism}

As each execution unit independently stores quantum circuits, it is crucial to manage the scheduling of execution modules efficiently and cost-effectively. We employ a process-based hierarchical triggering mechanism. By hierarchically storing the information of the quantum circuit in the server, controller and execution module, we achieve efficient top-down synchronous triggering.  For illustration, consider the two-level cascade shown on Fig.~\ref{fig:multiprocess}.

Initially, during compilation, the process ID is assigned to the quantum circuit to be executed, and the circuit is then converted into control programs for discrete execution units. The root controller records the leaf controller mask corresponding to the process ID and triggers only the leaf controller that matches this mask. The leaf controller maintains a log of the execution module mask associated with the process ID. Upon receiving the trigger signal for this process from the root controller, the leaf controller sends the process ID and synchronous trigger signal only to the respective execution module. Similarly, the execution module records the execution unit mask corresponding to the process ID and triggers only the specified execution unit based on the trigger signal and process ID.

\subsection{Multiprocessing Based Quantum Process-level Parallelism}
\label{sec_QPLP}
\subsubsection{Multiprocessing Microarchitecture}

Timing control of a process can be achieved through the process-based hierarchical trigger mechanism. Further, we utilize a multiprocessing scheme to realize quantum process-level parallelism. To support $N$-process parallel execution, each controller is equipped with $N$ task control processor, and each drive module with $N$ process managers, as depicted on Fig.~\ref{fig:multiprocess}(b) and (c).
\name currently supports up to 32 processes and is designed for seamless scalability to 64 or even 128 processes. However, since small-scale qubit systems are rarely used, the software has been configured to implement only 5 processes. 

When a new task is received by the controller, it is assigned to the corresponding task control processor and executed independently. The triggering arbitration and data transfer modules are primarily utilized to resolve process conflicts and forward readout results and feedback data to the appropriate task control processor. The emitter sends trigger signals and feedback data to the corresponding execution modules based on the process IDs in the configuration lookup table.

Focusing on drive modules as examples of execution modules, the process manager has the authority to configure and activate all qubit drive units, which are crucial for the configuration and maintenance of various processes. The dispatcher relays feedback data and trigger signals to specific units within the mask according to the process ID. The control of qubits for each process is facilitated by the qubit drive unit, independently, as previously stated.

\subsubsection{Staggered Trigger Mechanism}

\begin{figure}[htbp]
\centering
\footnotesize
\centering
    \includegraphics[width=0.45\textwidth]{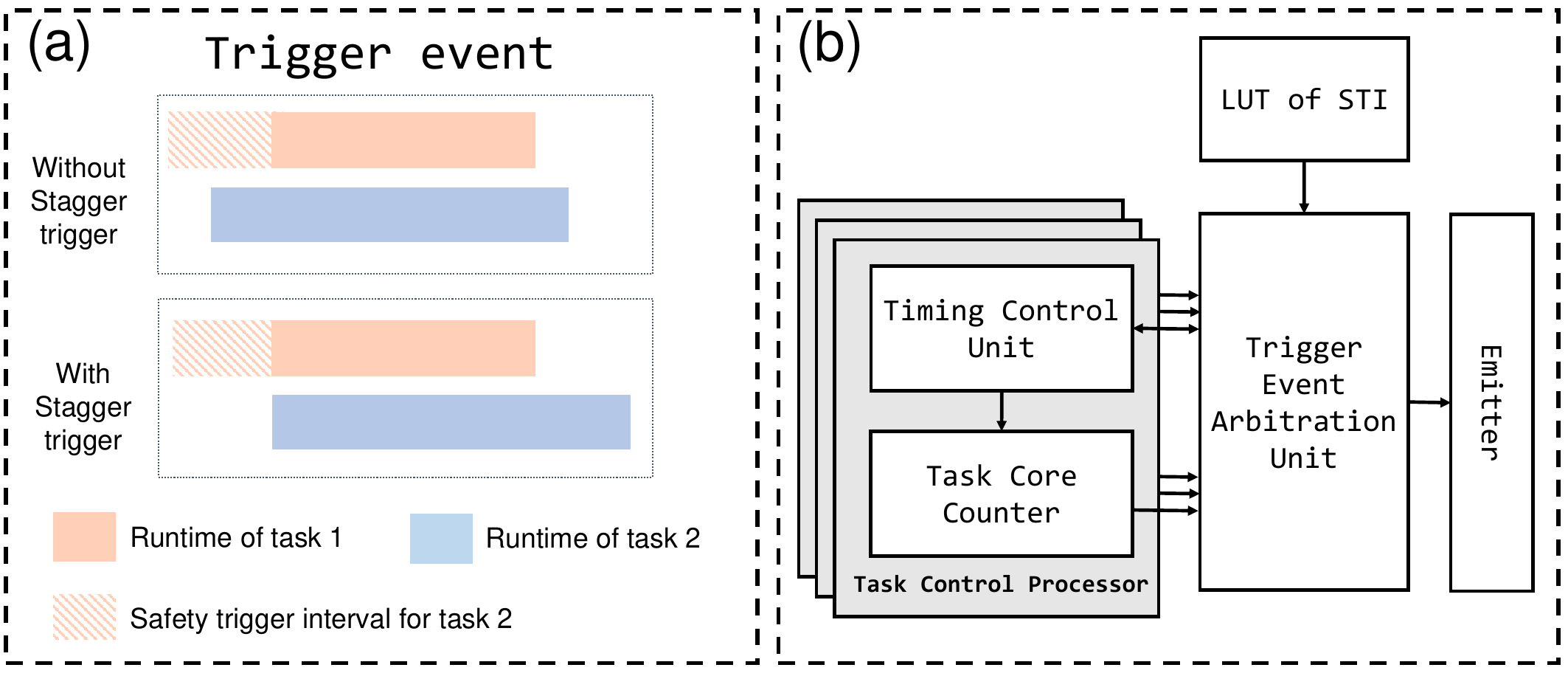}
\caption{(a) The upper half depicts the scenario where staggered trigger mechanism is not utilized, resulting in a triggering interval that is too short and may lead to inaccurate results due to crosstalk. The lower half illustrates the scenario where the staggered trigger scheme is employed. (b) The architecture for implementing the staggered trigger scheme. The safety trigger interval (STI) for each process is stored in a lookup table (LUT).}
\label{fig:trigger} 
\end{figure}

Crosstalk between qubit drive and readout lines is inevitable, potentially leading to conflicts between quantum processes and impacting the accuracy of results. When managing a single process, it is feasible to compensate for crosstalk during the compilation process, as the timing of quantum operations on each channel is precisely controlled. However, the timing becomes uncertain when two tasks are executed asynchronously in parallel. If control signals are applied concurrently in an area experiencing significant crosstalk, this can cause the signals received by the qubits to deviate from their calibrated values, potentially resulting in incorrect task outcomes.

To address this issue, we implement a staggered trigger mechanism to control the timing between process triggers. As shown in Fig.~\ref{fig:trigger}(a), when the interval between trigger events of two processes is too close, the trigger event arbitration unit of controller holds the trigger signal until the interval reaches the predefined safety threshold.  The specific implementation is depicted in Fig.~\ref{fig:trigger}(b). Each task control processor logs the time interval from the initial trigger using a task core counter and shares this information with the trigger event arbitration. To facilitate this, we introduce  \verb|start| as a parameter in the \verb|TRIGGER| instruction to denote whether a trigger is the initial one for a shot. The safety trigger interval (STI) for each process, specifying the minimum required timing intervals between processes, is stored in a lookup table. For trigger requests submitted by timing control units, the trigger event arbitration unit sends the trigger signal and responds to the timing control unit only when the interval between different processes meets the STI requirements.

\subsection{Feedback Control}
\label{feedback}

 This subsection outlines the implementation scheme for feedback control. Execution modules are designed to parse subsequent quantum circuits based on readout results. We have implemented \verb|BR| instructions for branch jumps. When a \verb|BR| instruction is parsed, the classical execution unit triggers an interrupt, and the parsing of the subsequent quantum circuit is paused until feedback data is received from the controller. 

For controllers, we employ a feedback interruption unit specifically tasked with handling feedback operations. It is important to note that only root controllers are responsible for making feedback decisions. Middle-layer controller boards serve primarily to transmit readout results and feedback data. The execution flow for controller is as follows:

{\bf Step 1.} Based on the input mask of the feedback entry table, it collects the readout results of the relevant qubits. 

{\bf Step 2.} After collecting the readout results, the feedback interruption unit, following the feedback decision unit, calculates feedback data of all qubits and the controller simultaneously.

{\bf Step 3.} Feedback data are sent to corresponding execution modules, and the interrupt is ended. One qubit corresponds to at least one qubit drive unit or qubit readout I/O unit.

{\bf Step 4.} To ensure timing synchronization after triggering, a \verb|TRIGGER 0 0| instruction is always issued immediately after the feedback instruction.

The feedback mechanism of HiMA supports the implementation of real-time quantum error correction, which is vital for advancing fault-tolerant quantum computing. The decoding circuits for quantum error correction codes are fundamentally similar to NISQ circuits. Moreover, error detection and recovery can be effectively accomplished through the aforementioned feedback mechanism by enabling the feedback decision unit to interpret the syndromes derived from measurements. This interpretation allows for the identification of the erroneous qubit and provides feedback data that instructs the execution module whether to jump to the branch program associated with error recovery. The interpretation of syndromes for Shor code and Steane code~\cite{nielsen2000QuantumComputationQuantum} is typically straightforward, usually achievable within one cycle, while the surface code~\cite{fowler2012SurfaceCodesPractical} demands a more substantial exertion to be accomplished within a practical timeframe~\cite{skoric2023ParallelWindowDecoding,zhang2024ClassicalArchitectureDigital}.

\section{Implementation}
\label{sec:implementation}

\begin{figure}[htbp]
\centering
\footnotesize
\centering
    \includegraphics[width=0.45\textwidth]{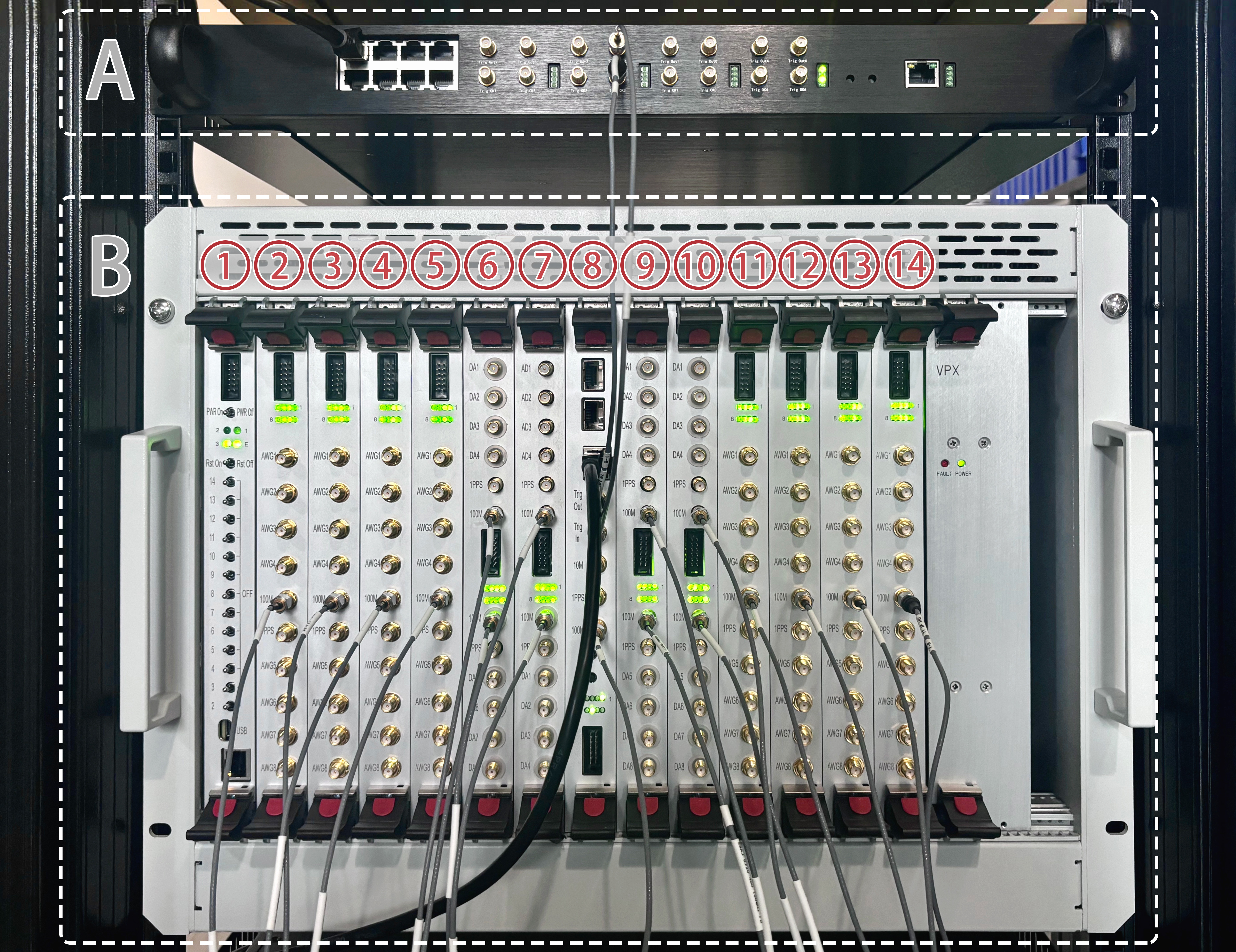}
\caption{Photograph for the implementation of a root controller (chassis A) and a qubit cluster control subsystem (chassis B). The first slot of the chassis B is for the power management board. Slots 2-5 and 11-14 are for Z drive boards. Slots 6, 9 and 10 are for XY drive boards. Slot 7 is for readout board. Slot 8 is for leaf controller board.}
\label{fig:implementation} 
\end{figure}

In this section, we elaborate on the implementation of \name. We develop a quantum control system that supports 72 tunable superconducting qubits, with a maximum task parallelism of 5, as detailed in the mechanisms discussed earlier. It serves for Origin Quantum Cloud Platform, which has been released publicly and has successfully completed a great number of user jobs. The target quantum processor has an average $T_1$ of \SI{14.51}{\micro\second} and an average $T_2$ of \SI{1.84}{\micro\second}. The single-qubit gates have a maximum and average fidelity of 99.85\% and 99.5\%, respectively. Fig.~\ref{fig:implementation} illustrates a root controller and a qubit cluster control subsystem.

\subsection{Implementation of execution boards}
Our system integrates five custom boards, each with specific roles and all centrally controlled by field programmable gate arrays (FPGAs). These boards include three execution boards, a root controller board and a leaf controller board.

The Z drive board features DACs with 1.2 GHz sample rate and 16-bit resolution, including 8 Z drive modules.
The XY drive board equipped with 6.4 GHz, 16-bit DACs. It produces 4-6 GHz signals directly using under-sampling technique for qubit manipulation.
Each board has 8 XY drive modules.
The readout board comprises 4 pairs of feedline output units and feedline input units, each pair independently supports 6-qubit readout.
It uses 16-bit, 6.4 GHz DACs and 11bit, 6.4 GHz analog-to-digital converters (ADCs).

All modules receive input clocks from a 100 MHz high-precision rubidium clock, distributed via a power distribution module, ensuring synchronized logic clocks across the system.

\subsection{Implementation of the system}
\label{sec:implemetation_system}
A QCCS, comprising 8 Z drive boards, 3 XY drive boards, 1 readout board and 1 leaf controller board, facilitates the management of 24 superconducting qubits with adjustable coupling.

To optimize the balance between efficiency and integration of the system, we utilize a VPX box for integration of the QCCS, and only the leaf controller board is connected to the server via a 10 Gigabit TCP link for transmitting and receiving packets. The leaf controller board forwards the data through a high-speed serial bus on the backplane, utilizing the AURORA 64b66b protocol. Feedback data, trigger signals and reset signals are conveyed using low voltage differential signaling (LVDS) through the backplane, with a maximum rate of 1.2 Gbps for a single pair of signals.

            
The scalability of the system primarily relies on synchronization and data transfer between the subsystems, which is facilitated through the root controller board, shown in the chassis A of Fig.~\ref{fig:implementation}. Due to the considerable distances between the root controller boards and the subsystems, we employ higher-speed LVDS communication between the root controller board and the leaf controller board for efficient transmission of control signals and feedback data.

Currently, the root controller board supports the control of up to 8 QCCSs, supports a maximum of 192 tunable superconducting qubits and 768 fixed-frequency qubits according to the quantum device of IBM. By further introducing three-layer cascading, the system's capacity can be expanded to support up to 6144 qubits.

\section{Evaluation}
\label{sec:evaluation}
In this section, we show how \multithreadingshort leads to speedups in terms of QPU Load Average and Circuit Layer Operations Per Second (CLOPS).
Furthermore, we present a novel application of quantum multiprocessing: the characterization of crosstalk between qubits.
\subsection{QPU Load Average}
We set 1024 shots per quantum circuit as our standard configuration throughout experiments. Each shot has a fixed \SI{100}{\micro\second} duration to ensure the qubit relaxation. To simplify the analysis, we assumed any qubit can be manipulated simultaneously.

\begin{figure}[htb]
	\centering
	{\includegraphics[width=\linewidth]{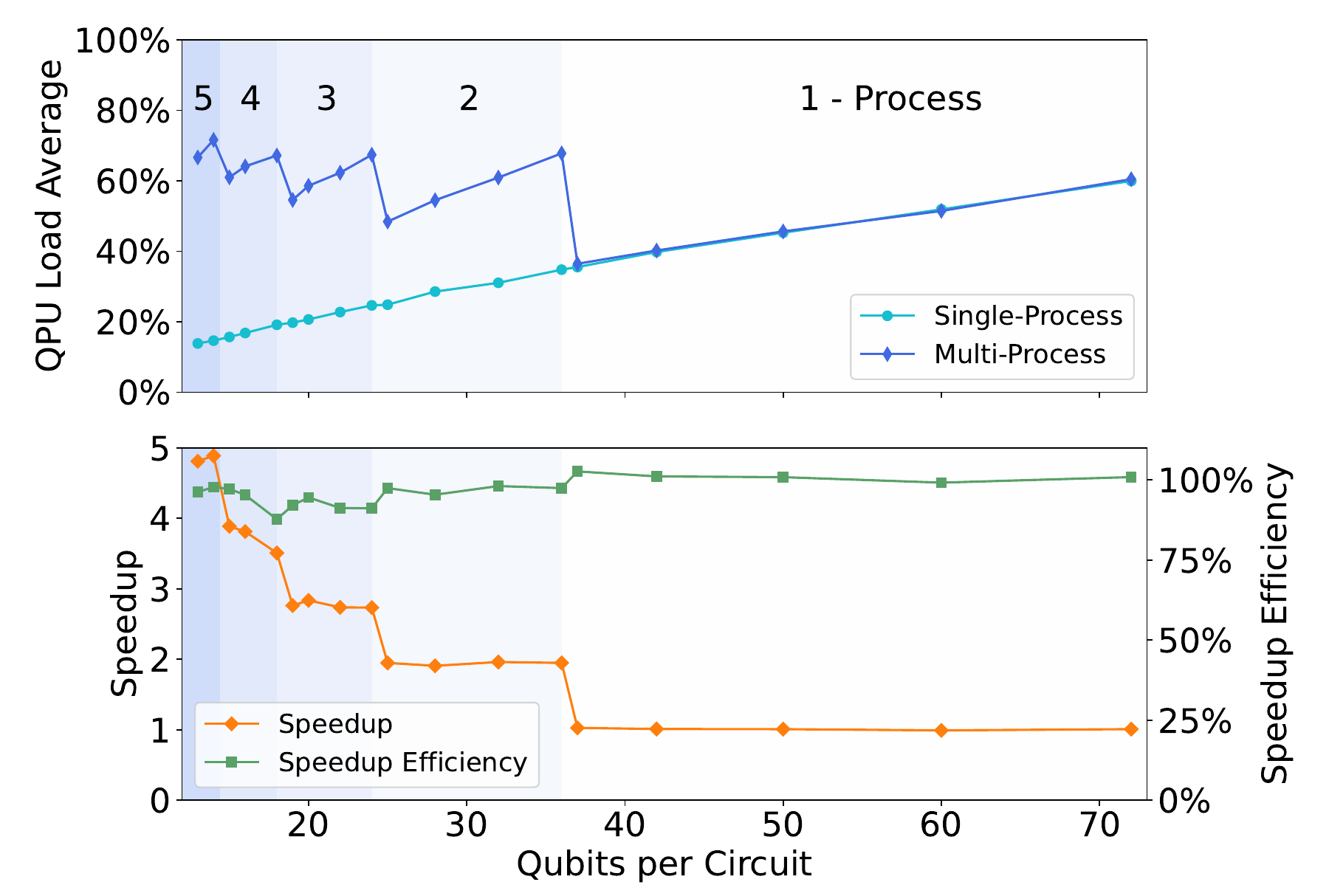}}
	\caption{Results of benchmark tests under varying \qubitsusage. The number of processes refers to the maximum number of cores that can be run concurrently at a given \qubitsusage. Due to fluctuations in the measured time, the efficiency calculated at the end may slightly exceed 100\%.}
	\label{fig:evaluation1} 
\end{figure}

We use \metricsshort, \speedup and \speedupefficiency as the metric for \multithreadingshort exploitation, \speedup is defined as
$$
\mathrm{Speedup}=\frac{t_\mathrm{total,Single\mbox{-}process}}{t_\mathrm{total,Multi\mbox{-}process}},
$$
and \speedupefficiency is defined as
$$
\mathrm{Speedup\,Efficiency}=\frac{\mathrm{Speedup}}{\mathrm{Number\,of\,Processes}},
$$
where Number of Processes refers to the maximum number of processes that can be run concurrently at particular \qubitsusage. For instance, when the \qubitsusage is 15, four processes can be executed simultaneously. The \speedupefficiency quantifies the ratio between the actual and ideal speedups, effectively gauging the overall overhead of multiprocessing scheduling.

In the evaluation, we varied the \qubitsusage, which is the percentage of the total qubits in the QPU used by a given circuit.
The results are shown in Fig.~\ref{fig:evaluation1}, number of processes is represented by background color, splitting the panel into five zones. In each process zone, the positive correlation between the \metricsshort and \qubitsusage is evident, as more \qubitsusage directly translate to fewer idle qubits. We found that multiprocessing can significantly improve \metricsshort in the case of fewer qubits ($<18$), increasing it from 16.03\% to \textbf{66.11\%}.
We also observed that the \speedup under all five process zones approaches the ideal speedup, producing a high \speedupefficiency. Overall, the 5-core system achieves a maximum \speedup of {\bf 4.89$\times$} and for cases with more than one process, we were able to achieve an average speedup efficiency of {\bf 94.48\%}.

\subsection{Circuit Layer Operations Per Second}
\label{sec:evaluation_CLOPS}

\begin{table}[ht]
\caption{CLOPS Test Results}
\centering
\label{tab:clops}
\begin{tabular}{lccc}
\toprule
 & \multicolumn{1}{c}{$\log_2\mathrm{QV}$} & \multicolumn{1}{c}{CLOPS} & \multicolumn{1}{c}{\begin{tabular}[c]{@{}c@{}}Efficiency factor\\ CLOPS/D\end{tabular}}\\
\midrule
IBM & 9 & 15,000 & 1,666.7 \\
Rigetti & Not released & 892 & $\leq 892$ \\
\name (this paper) & 5 & 12,304 & 2,460.8\\
\bottomrule
\end{tabular}
\end{table}

The results of the CLOPS test are presented in Table~\ref{tab:clops}, where we compare our system with the publicly available data from IBM~\cite{IBM} and Rigetti~\cite{Rigetti}. All CLOPS tests were performed using the aforementioned cloud platform. The results demonstrate that \name exhibits higher efficiency compared to all known platforms. We conclude the improvement in the efficiency is mainly due to our architecture design, which provides a better issue rate.

Moreover, the multi-processing provides significant improvement for CLOPS tests. Multi-process CLOPS are tested on 2 to 5 independent regions of the quantum chip, with results shown in Table~\ref{tab:clops2}. Results show the 5-process CLOPS reaches 43,680, an almost 4-fold improvement over the 1-process case. Note that the multi-process scheduling is automatically conducted through the cloud platform, and is transparent to common users. This implies a general efficiency advantage on \name as well as our quantum cloud platform.

\begin{table}[ht]
\caption{CLOPS for multi-processing}
\centering
\label{tab:clops2}
\begin{tabular}{cccccc}
\toprule
No. of Proc. & 1 & 2 & 3 & 4 & 5\\
\midrule
Total CLOPS & 12,305 & 23,218 & 29,778 & 38,024 & 43,680 \\
CLOPS per Proc. & 12,305 & 11,609 & 9,926 & 9,506 & 8,736 \\

\bottomrule
\end{tabular}
\end{table}
\subsection{Characterizing Crosstalk between Qubits via Quantum MultiProcessing}
In this subsection, we introduce a novel application of quantum multiprocessing for characterizing crosstalk between qubits.

\begin{figure}[htb]
	\centering
	\includegraphics[width=\linewidth]{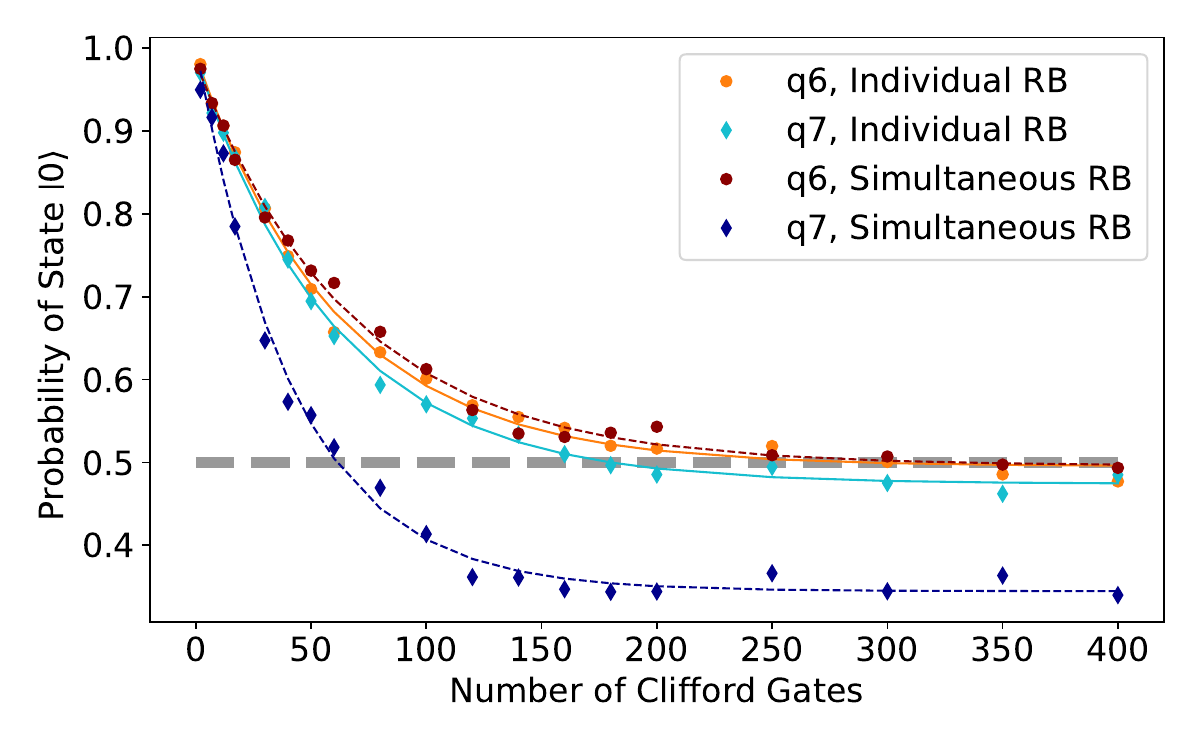} \\
	\caption{Experiments for individual RB and simultaneous RB on two qubits (q6 and q7). One Clifford gate corresponds to an average of 1.875 single-qubit gates of \SI{30}{\nano\second} duration. Symbols represent the average of experimental data, and solid (dashed) curves are the fitting results. Note that the convergence value of q7 for simultaneous RB is diminished due to $\ket2$ state leakage introduced by crosstalk when controlling q6.}
	\label{fig:experiment} 
\end{figure}

Individual~\cite{kayanuma2008CoherentDestructionTunneling,proctor2017WhatRandomizedBenchmarking} and simultaneous~\cite{gambetta2012CharacterizationAddressabilitySimultaneous} randomized benchmarking (RB) experiments were conducted on two neighboring qubits (q6 and q7) on our 72-qubit QPU. By flexibly controlling the trigger interval between two RB tasks, we observe a continuous change of the behavior of one qubit. The results are shown in Table.~\ref{tab:experiment} and Fig.~\ref{fig:experiment}.

\renewcommand{\arraystretch}{1}
\begin{table}[htb]
	\centering
	\caption{Convergence value of q6 and q7 in RB experiment under different trigger interval.}
	\label{tab:experiment}
	\begin{tabular}{ccc}
		\toprule
		Trigger Interval &\makecell{Convergence\\Value of q6} &\makecell{Convergence\\Value of q7}\\
		\midrule
		Simultaneous           &$0.49\pm0.03$ &$0.34\pm0.03$\\
		\SI{5}{\micro\second}  &$0.52\pm0.03$ &$0.38\pm0.02$\\
		\SI{10}{\micro\second} &$0.50\pm0.03$ &$0.46\pm0.03$\\
		\SI{15}{\micro\second} &$0.49\pm0.02$ &$0.48\pm0.03$\\
		\SI{20}{\micro\second} &$0.49\pm0.04$ &$0.50\pm0.03$\\
		Individual             &$0.48\pm0.04$ &$0.49\pm0.04$\\
		\bottomrule
	\end{tabular}
\end{table}

Individual RB experiments reveal average single-qubit gate fidelities of 99.57\% and 99.56\% for qubits q6 and q7, respectively. However, simultaneous RB experiments yield fidelities of 99.61\% for qubit q6 and 99.38\% for qubit q7. 
The reason for this is that the energy gap between the states $\ket1$ and $\ket2$ of q7 has been tuned to closely match the energy gap between the states $\ket0$ and $\ket1$ of q6. Consequently, during simultaneous RB, operations on q6 are more likely to cause q7 to leak to the state $\ket2$ through crosstalk. This results in a reduction of q7's fidelity while q6 remains largely unaffected.

In our design, the trigger interval can be adjusted to perform multiple tasks simultaneously on different cores at staggered intervals to mitigate the effect of cross-talk. Thus, we further performed RB experiments with different trigger intervals on the above two qubits.
Since the negligible impact of crosstalk on fidelity and the substantial fluctuation in fidelity due to qubit instability, we opt for the convergence value (CV), a more stable metric, to assess crosstalk effects across varying trigger intervals.
As shown in Fig.~\ref{fig:experiment}, the CV is around 0.5 under normal conditions, and the CV will decrease when affected by crosstalk. This is because leakage to the state $\ket2$ makes the state discrimination result more biased toward the state $\ket1$.
As shown in Table~\ref{tab:experiment}, the CV of q7 increases proportionally to the trigger interval, nearing the result of independent execution when the trigger interval is sufficiently large. Meanwhile, the CV of q6 remains stable at around 0.5, consistent with expectations.

These results show that \name can flexibly run multiple independent experiments in parallel, while ensuring their respective timing requirements.

\section{Related Work}
\label{sec:related_work}

\textbf{Multi-Programming:} A multi-programming method proposed in~\cite{das2019CaseMultiProgrammingQuantum} facilitates the concurrent execution of multiple tasks in software. This synchronous parallel approach merges tasks that can be executed simultaneously into a batch during compilation to enhance efficiency. However, it often leads to inefficiency and inflexibility due to several inherent limitations. Firstly, integrating various tasks into a single quantum program means that both the execution time and the number of sampling iterations are dictated by the largest task. This arrangement can lead to the underutilization of the capabilities of the quantum device. Secondly, because this parallelism is implemented at the software level, it is not feasible to insert new executable tasks during the execution of other quantum programs until the next batch can be organized and executed, resulting in lower utilization rates of qubit resources. Lastly, the diverse nature of tasks running on quantum devices, ranging from qubit calibration to complex quantum algorithms, necessitates sophisticated scheduling algorithms. This requirement introduces additional time overhead and diminishes the overall efficiency and agility of the quantum computing system. In contrast, we introduce a novel multiprocessing microarchitecture that supports asynchronous parallel execution. Our approach enhances flexibility and efficiency by enabling independent compilation and execution of multiple tasks.

\textbf{Comparison of Microarchitectures:} IBM proposes a hierarchical architecture in Ref.~\cite{gupta2024encoding}; however, it lacks support for quantum process-level parallelism and does not provide a concrete implementation scheme. QuMA~\cite{fu2017ExperimentalMicroarchitectureSuperconducting} initially employs a centralized architecture, limited by issue rate constraints. To address this, QuMA$\underline{~}$v2~\cite{fu2019EQASMExecutableQuantum} integrates Single-Operation-Multiple-Qubit and Very-Long-Instrction-Word, while QuAPE~\cite{zhang2021ExploitingDifferentLevels}  introduces a multiprocessor for quantum superscalar expansion. However, these centralized architectures still face scalability challenges due to resource and bandwidth limitations. To overcome these issues, a classical architecture with Single-Instruction-Multiple-Data and broadcasting mechanisms is proposed in~\cite{zhang2024ClassicalArchitectureDigital}. This approach reduces storage and communication pressure by storing channel information at different abstraction levels, thus enhancing scalability. Nevertheless, issue rate and efficiency remain bottlenecks for complex quantum circuits, such as those used in parametric quantum computation~\cite{Funcke2021dimensional}. In this paper, HiMA adopts a hierarchical architecture to separate quantum gate information from the root controller, ensuring that resolution efficiency does not increase with the complexity of the quantum circuit. Moreover, our microarchitecture supports quantum process-level parallelism, significantly improving efficiency and qubit utilization.

\section{Conclusion}

In conclusion, we introduce \name, a hierarchical quantum microarchitecture that addresses the pressing challenge of scalability in quantum computing control systems. Through a modular approach, precise timing control mechanisms, and asynchronous measurement techniques, \name effectively achieves quantum process-level parallelism at the granularity of individual qubits. 

As a result, we deploy \name as a control system for a 72-qubit superconducting QPU, used for quantum cloud computing. It is readily extendable to accommodate up to 768 fixed-frequency qubits. By further introducing three-layer cascading, the system's capacity can be expanded to support up to 6144 qubits. 
In benchmarking tests, we demonstrate the inherent speedup achieved through multiprocessing with \name, showcasing significant \metrics (\metricsshort) improvement (from 16.03\% to 66.11\%) at low \qubitsusage and achieves up to a 4.89× speedup and an average speedup efficiency of 94.48\% under a 5-process parallel configuration. 
The CLOPS of our test system can reach up to 43,680, measured through the cloud platform, which accounts for real-world latency. 
Moreover, in randomized benchmarking experiments, \name employs adjustable staggered triggering techniques to mitigate crosstalk, confirming \name's reliability and flexibility of asynchronous parallelism of processes. By enabling efficient and scalable quantum computing, \name paves the way for more complex quantum algorithms and applications, promising to unlock new possibilities in quantum research and technology.

\section*{Acknowledgement}
This work is supported by National Key Research and Development Program of China (Grant No. 2023YFB4502500).

\bibliographystyle{unsrt}
\bibliography{quantum,manual}

\end{document}